\documentclass{PoS}

\title{Recent results on $N^*$ spectroscopy with ANL-Osaka dynamical coupled-channels approach}

\ShortTitle{Recent results on $N^*$ spectroscopy with ANL-Osaka dynamical coupled-channels approach}

\author{\speaker{H.~Kamano}\\
        Research Center for Nuclear Physics (RCNP), Osaka University, Ibaraki, Osaka 567-0047, Japan\\
        E-mail: \email{kamano@rcnp.osaka-u.ac.jp}}
\author{S.X.~Nakamura\\
        Department of Physics, Osaka University, Toyonaka, Osaka 560-0043, Japan\\
        E-mail: \email{sxnakamura@gmail.com}}
\author{T.-S.~H.~Lee\\
        Physics Division, Argonne National Laboratory, Argonne, Illinois 60439, USA\\
        E-mail: \email{lee@phy.anl.gov}}
\author{T.~Sato\\
        Department of Physics, Osaka University, Toyonaka, Osaka 560-0043, Japan\\
        E-mail: \email{tsato@phys.sci.osaka-u.ac.jp}}

\abstract{
We report our recent effort on the extraction of 
$N^*$- and $\Delta^*$-resonance parameters (pole masses, coupling constants, etc.) 
through the comprehensive analysis of pion- and photon-induced 
$\pi N$, $\eta N$, $K\Lambda$, and $K\Sigma$ production reactions.
The analysis was performed with the ANL-Osaka dynamical coupled-channels approach,
which satisfies the multichannel unitarity of the S-matrix
in the channel space spanned by $\pi N$, $\eta N$, $K\Lambda$, $K\Sigma$, 
and also three-body $\pi \pi N$ 
that contains $\rho N$, $\pi \Delta$, and $\sigma N$ resonant components.
Ongoing projects and future plans are also discussed.
}

\FullConference{XV International Conference on Hadron Spectroscopy-Hadron 2013\\
		4-8 November 2013\\
		Nara, Japan }

\begin{document}

\section{Introduction}

An understanding of the mass spectrum and structure of the excited nucleons ($N^*$) 
is a fundamental challenge in the hadron physics. 
So far, a number of static hadron models such as constituent quark models 
have been developed to study the $N^*$ spectrum and form factors,
in which the excited states are treated as stable particles.
However, in reality the $N^*$ states couple strongly to the meson-baryon continuum 
states and can exist only as unstable resonances in the $\pi N$ and $\gamma N$ reactions. 
Such a strong coupling to the continuum states would affect significantly 
the properties of the $N^*$ states and should not be ignored in extracting the $N^*$ resonance 
parameters (pole masses, coupling constants, etc) from the data 
and giving physical interpretations to those parameters.

\begin{figure}[b]
\begin{center}
\includegraphics[width=0.85\textwidth,clip]{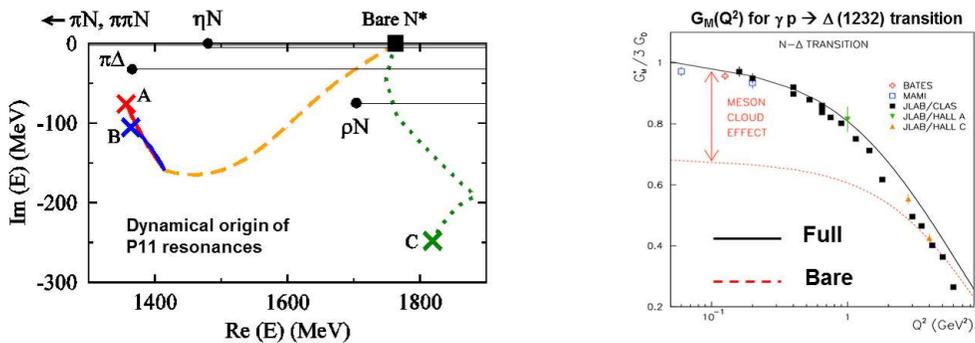}
\caption{\label{fig1}
(Left panel) Dynamical origin of $P_{11}$ nucleon resonances~\cite{sjklms10}.
(Right panel) Meson-cloud effect on the $N$-$\Delta$ $M1$ transition form factor~\cite{sl2}.
}
\end{center}
\end{figure}

Over the past years, we have been working on the spectroscopy of 
the $N^*$ states in terms of a reaction theory.
From this study it has become clear that the multichannel 
reaction dynamics, which is responsible for turning the $N^*$ states into unstable 
resonances, indeed plays a crucial role for understanding properties of the $N^*$ states 
as resonances both qualitatively and quantitatively.
For example, it was found in our early analysis~\cite{sjklms10}
that the multichannel reaction dynamics can generate many physical resonances 
from just a single ``bare state'' that can be associated with a baryon defined 
within static hadron models [Fig.~\ref{fig1} (left panel)].
This result implies that within the multichannel reactions, a naive one-to-one 
correspondence between static baryons and physical resonances does not exist in general.
Also, the $N\to\Delta(1232)$ $M1$ transition form factor extracted, e.g., in Ref.~\cite{sl2}
[Fig.~\ref{fig1} (right panel)] 
demonstrates the importance of the meson cloud effect, which originates from 
the reaction dynamics, for a quantitative understanding of the $N^*$ form factors.

The pion- and photon-induced meson production reactions off a nucleon 
is known as the most useful reactions for studying $N^*$ resonances.
In fact, in those reactions the $N^*$ resonances appear as direct s-channel processes 
and most of the information on the resonances can be obtained by analyzing 
the reaction cross sections for various final states.
Such an analysis is feasible because a huge amount of the high precision data 
of meson photoproductions has been measured at facilities such as JLab and CBELSA 
and is available for the detailed partial wave analysis.

In this contribution, we report our recent results on
the comprehensive analysis of pion- and photon-induced meson-production reactions
with the ANL-Osaka dynamical coupled-channels (DCC) approach.

\section{ANL-Osaka DCC analysis}

\begin{figure}[t]
\begin{center}
\includegraphics[width=0.55\textwidth,clip]{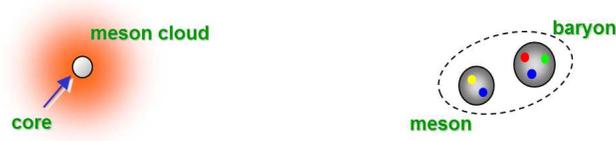}
\caption{\label{fig2}
Schematic pictures of $N^*$ resonances.
(Left)
The bare state (``core'') surrounded by meson clouds, where the core
would correspond to a baryon defined within static hadron models.
(Right) Meson-baryon molecule-like resonance.
}
\end{center}
\end{figure}

Within the ANL-Osaka DCC approach~\cite{msl,dcc8},
the partial wave amplitudes $T^{IJ(LS)}$ are obtained by solving
the following coupled-channels integral equation:
\begin{equation}
T_{b,a}^{IJ(LS)}(p_b,p_a;E) =
V_{b,a}^{IJ(LS)}(p_b,p_a;E) +\sum_c \int_C q^2 dq
V_{b,c}^{IJ(LS)}(p_b,q;E) G_c(q;E) T_{c,a}^{IJ(LS)}(q,p_a;E),
\label{eq1}
\end{equation}
\begin{equation}
V_{b,a}^{IJ(LS)}(p_b,p_a;E) = v_{b,a}^{IJ(LS)}(p_b,p_a;E) 
+ \sum_{{\rm all}~N^*_0~{\rm for~given~} IJ(LS)} \frac{\Gamma_{b,N^*_0}(p_b) \Gamma_{N^*_0,a}(p_a)}{E-m_{N^*_0}}.
\label{eq2}
\end{equation}
Here the subscripts denote the reaction channels 
$a,b,c = \pi N, \eta N, K\Sigma, K\Lambda, \pi \Delta, \rho N, \sigma N, \gamma N$;
the potential $V^{IJ(LS)}$ consists of the so-called ``non-resonant'' process
($v_{b,a}^{IJ(LS)}$), in which only mesons and baryons belonging to the grand states of each SU(3)
flavor multiplet are included, and the s-channel process of ``bare'' $N^*$ states ($N^*_0$); 
$G_c$ is the Green's function for the channel $c$;
and the momentum integral path $C$ is chosen appropriately in the complex
$q$-plane within the range $0 \leq |q| < \infty$.
The Green's functions of the quasi-two-body channels $\pi \Delta$, $\rho N$, and $\sigma N$
have a self energy arising from the 
$\Delta \to \pi N$, $\rho \to \pi \pi$, and $\sigma \to \pi \pi$ decays, respectively.
This ensures that the resulting partial wave amplitudes have appropriate branch points
for the $\pi \Delta$, $\rho N$, $\sigma N$ as well as three-body $\pi \pi N$ channels
in the complex $E$-plane.
Our approach satisfies the coupled-channels unitarity of all major meson-baryon channels 
including the 3-body $\pi\pi N$ for the considered energy region,
and the momentum integral in Eq.~(\ref{eq1}) ensures the proper treatment of 
the off-shell dynamics, which are not possible within the on-shell K-matrix approaches.
Furthermore, the parametrization of the potential [Eq. (\ref{eq2})]
allows us to treat the two pictures of the baryons, i.e., ``core + meson clouds'' and 
``meson-baryon molecules'' (Fig.~\ref{fig2}), on the same footing.

\begin{figure}[t]
\begin{center}
\includegraphics[width=0.8\textwidth,clip]{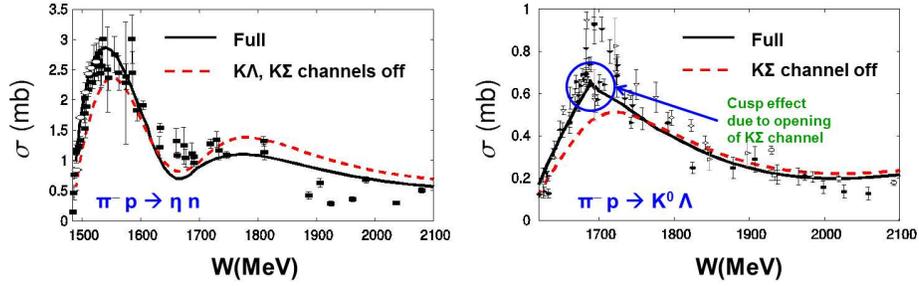}
\caption{\label{fig3}
Coupled-channels effect on the reaction cross sections.
(Left panel) $\pi^- p \to \eta n$ reaction total cross section;
(Right panel) $\pi^- p \to K^0\Lambda$ reaction total cross section.
The black solid curves are the full results computed with our 8-channel DCC model~\cite{dcc8},
while the red dashed curves are the results in which coupling to the $KY$ channels
is turned off.
}
\end{center}
\end{figure}

Recently, we have completed~\cite{dcc8} a fully combined analysis of 
the $\pi N \to \pi N, \eta N, K \Lambda, K \Sigma$ and
$\gamma N \to \pi N, \eta N, K \Lambda, K \Sigma$ reactions
and determined our model parameters contained in the potential~(\ref{eq2}),
where the 8 reaction channels 
($\pi N, \eta N, K\Sigma, K\Lambda, \pi \Delta, \rho N, \sigma N, \gamma N$)
are taken into account in solving scattering equations.
The data used in this analysis are the SAID energy-independent solutions~\cite{saiddb} 
up to $W=2.3$ GeV for the $\pi N$ scattering, while 
the actual unpolarized and polarized observables up to $W=2.1$ GeV
are used for the other reactions.
This results in fitting more than 22,000 data points.
The total cross sections of $\pi^- p \to \eta n$ and $\pi^- p \to K^0 \Lambda$
computed with our constructed 8-channel DCC model 
are presented in Fig.~\ref{fig3} (black solid curves),
showing a good agreement with the available data 
(note that the total cross section data are not included in our analysis).
In the same figures, we also present the results (red dashed curves) in which 
coupling to the $KY$ channels are turned off to demonstrate
the coupled-channels effect of the $KY$ channels on the observables.
One can see the sizable coupled-channels effect from the $KY$ channels.
In particular, it is interesting to see that in our model the sharp peak of 
the $\pi^- p \to K^0 \Lambda$ total cross section around $W=1.7$ GeV 
is explained as the cusp effect due to opening of the $K\Sigma$ channel.

\begin{figure}[t]
\begin{center}
\includegraphics[width=0.8\textwidth,clip]{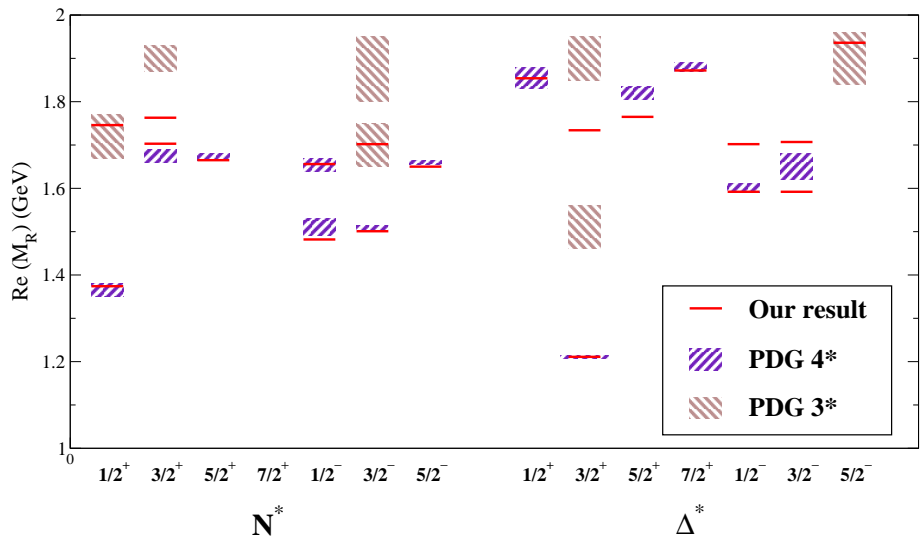}
\caption{\label{fig4}
$N^*$ mass spectrum extracted via the 8-channel DCC analysis~\cite{dcc8}.
The real parts of the complex $N^*$ pole mass $M_R$ are plotted as red lines. 
As a comparison, the spectrum of four- and three-star $N^*$ resonances assigned by PDG~\cite{pdg12}
is also plotted as shaded rectangles that represent the range of ${\rm Re}(M_R)$
obtained from previous analyses selected by PDG.
}
\end{center}
\end{figure}

The $N^*$ mass spectrum extracted from our analysis is presented in Fig.~\ref{fig4}.
Here we have plotted only $N^*$ resonances with 
${\rm Re}(M_R) < 2$ GeV and ${\rm Im}(M_R) < 0.2$ GeV, which are expected to be extracted
with confidence within our model.
Here we would like to comment on the second $\Delta(J^P = 3/2^+)$ resonance,
the Roper-like state of the $\Delta$ baryons.
At present, a sizable ambiguity exists for the parameters of this resonance,
while the properties of the first $\Delta(3/2^+)$ resonance
[$\Delta(1232)$ in the notation of PDG~\cite{pdg12}] have been determined very accurately.
This is because the $P_{33}$ $\pi N$ partial wave amplitude
is dominated by $\Delta(1232)$ at low energies, while the second $\Delta(3/2^+)$
has little contribution to it.
It is thus very important to find out reactions that are sensitive to 
this Roper-like state of $\Delta$ in shedding light on the nature of it.

\section{Ongoing projects and future plans}

As an immediate application of our 8-channel DCC model, we plan to extract 
the $N\to N^*$ electromagnetic transition form factors 
up to $Q^2=6$ (GeV/c)$^2$ for $N^*$ resonances with 
${\rm Re}(M_R) \leq 1.6$ GeV by analyzing all available data of the $p(e,e' \pi)N$ reaction from CLAS.
In parallel with this, we keep extending our model. 
We first plan to extend our model by adding the $\omega N$ channel and 
performing the 9-channel analysis including the $\omega N$ production data.
After this completes, we will further improve our model by including the double pion production data.
By this stage, we hope to have the extensive data of $\pi N \to \pi \pi N$ that 
are planned to be measured at J-PARC~\cite{e45,kam13}, so that we can make a
comprehensive analysis including both the $\pi N \to \pi \pi N$ and $\gamma N\to \pi \pi N$ data.
The extension of our model to deuteron-target reactions is also planned for 
the purpose of extracting electromagnetic transition form factors associated with neutron.
Furthermore, applications of our DCC approach to the $Y^*$ spectroscopy, 
the meson spectroscopy~\cite{meson1,meson2}, 
and the neutrino reactions~\cite{neutrino} are also underway.
Developing the DCC model for neutrino-induced reactions are 
very important for the comprehensive understanding of the neutrino-nucleon/nucleus interactions
in the energy region relevant to the neutrino-oscillation experiments~\cite{nuint}.

This work was supported by the JSPS KAKENHI Grant 
No.~25800149 (H.K.) and No.~24540273 (T.S.),
and by the U.S. Department of Energy, Office of Nuclear Physics Division,
under Contract No. DE-AC02-06CH11357.
H.K. acknowledges the support from the HPCI Strategic Program
(Field 5 ``The Origin of Matter and the Universe'') of MEXT of Japan.


\begin{thebibliography}{99}
\bibitem{sjklms10}
N.~Suzuki, B.~Juli\'a-D\'iaz, H.~Kamano, T.-S.~H.~Lee, A.~Matsuyama, and T.~Sato,
Phys. Rev. Lett. {\bf 104}, 042302 (2010).

\bibitem{sl2}
B.~Juli\'a-D\'iaz, T.-S.~H.~Lee, T.~Sato, and C. Smith,
Phys. Rev. C {\bf 75}, 015205 (2007).

\bibitem{msl}
A.~Matsuyama, T.~Sato, and T.-S.~H.~Lee,
Phys. Rep. {\bf 439}, 193 (2007).

\bibitem{dcc8} 
H.~Kamano, S.X.~Nakamura, T.-S.~H.~Lee, and T.~Sato,
Phys. Rev. C {\bf 88}, 035209 (2013).

\bibitem{pdg12}
J.~Beringer {\it et al.} (Particle Data Group), 
Phys. Rev. D {\bf 86}, 010001 (2012).

\bibitem{saiddb}
CNS Data Analysis Center, George Washington University,
http://gwdac.phys.gwu.edu.

\bibitem{e45}
K.~Hicks and H.~Sako {\it et al.} (J-PARC E45 experiment),
http://j-parc.jp/researcher/Hadron/en/pac\_1207/pdf/P45\_2012-3.pdf.

\bibitem{kam13}
H.~Kamano,
Phys. Rev. C {\bf 88}, 045203 (2013).

\bibitem{meson1}
H.~Kamano, S.~X.~Nakamura, T.-S.~H.~Lee, and T.~Sato,
Phys. Rev. D {\bf 84}, 114019 (2011).

\bibitem{meson2}
S.~X.~Nakamura, H.~Kamano, T.-S.~H.~Lee, and T.~Sato,
Phys. Rev. D {\bf 86}, 114012 (2012).

\bibitem{neutrino}
H.~Kamano, S.~X.~Nakamura, T.-S.~H.~Lee, and T.~Sato,
Phys. Rev. D {\bf 86}, 097503 (2012).

\bibitem{nuint}
http://nuint.kek.jp/index\_e.html; 
S.~X.~Nakamura, Y.~Hayato, M.~Hirai, H.~Kamano, S.~Kumano, M.~Sakuda, K.~Saito, and T.~Sato,
arXiv:1303.6032.

\end{thebibliography}
\end{document}